# Horizon-T Experiment Calibrations – MIP Signal from Scintillator and Glass Detectors

D. Beznosko [1b], A. Iakovlev [b], A. Duspayev [b], M. I. Vildanova [a], T. Uakhitov [b], M. Yessenov [b], V.V. Zhukov [a]

*Abstract*

Horizon-T, a modern Extensive Air Showers (EAS) detector system, is constructed at Tien Shan high-altitude Science Station of Lebedev Physical Institute of the Russian Academy of Sciences at approximately 3340 meters above the sea level in order to study in the energy range above $10^{16}$ eV coming from a wide range of zenith angles (0° - 85°).

The detector includes eight charged particle detection points and a Vavilov-Cherenkov radiation detector. Each charged particle detector response is calibrated using single MIP (minimally ionizing particle) signal. The details of this calibration are provided in this article. This note is valid for data before March 2017 and will not be updated following any detector calibration and configuration changes as a large upgrade has been implemented.

## 1. Detector System Brief Description

Tien Shan high-altitude Science Station, a part of Lebedev Physical Institute of the Russian Academy of Sciences, is located 32 km from the city of Almaty at the altitude of ~3340 meters above the sea level. “Horizon-T” (HT) detector system [1] [2] is constructed to study space-time distribution of the charged particles in EAS disk and Vavilov-Cherenkov radiation from it with parent particle of energies higher than $10^{16}$ eV coming from a range of zenith angels (0°-85°). The novel method of using time information from pulse shape in each detector allows for the analysis of EAS with core falling outside of the detector system bounds.

HT consists of eight charged particles detection points and a detection of the Vavilov-Cherenkov radiation (VCD). The relative coordinates of every station and distances from station 1 are presented in Table 1. The aerial view of HT with detection points marked is in Figure 1.

All detection points and VCD connect to Data Acquisition system (DAQ) via cables. The cable calibration procedure and results are given is [3]. There are plans to upgrade the detector system to cable-less [4]. In order to study the spatial distribution of charged particles in EAS disk, an accurate calibration to a single particle response is required.

## 2. Near and Far Periphery Detectors

The near periphery combines the detection points 1, 4, 5, 6 and 7. Each of these detection points has one scintillator detector (SD) and one glass detector (GD). Both types have a pyramid shape with height equal to the side. Each SD is based on polystyrene square-shaped cast [5] scintillator [6] that has 1 $m^2$ area and is 5 cm thick. Each GD is based on 50 cm x 50 cm x 3 cm optical glass that is painted white with $TiO_2$ at the bottom side [7]. Both have the PMT above the scintillator/glass center. All near detectors use Hamamatsu [8] R7723 PMT assembly. Only the near periphery is equipped with GD: the fast pulse they produce gets widened by the longer cable thus diminishing the usefulness of the GL. Center temporarily has a second GL with a Hamamatsu H6527 PMT and its calibration is also included in the table.

---

[1] dmitriy.beznosko@nu.edu.kz (also dima@dozory.us)
[a] P. N. Lebedev Physical Institute of the Russian Academy of Sciences, Moscow, Russia
[b] Physics Department, Nazarbayev University, Astana, Kazakhstan

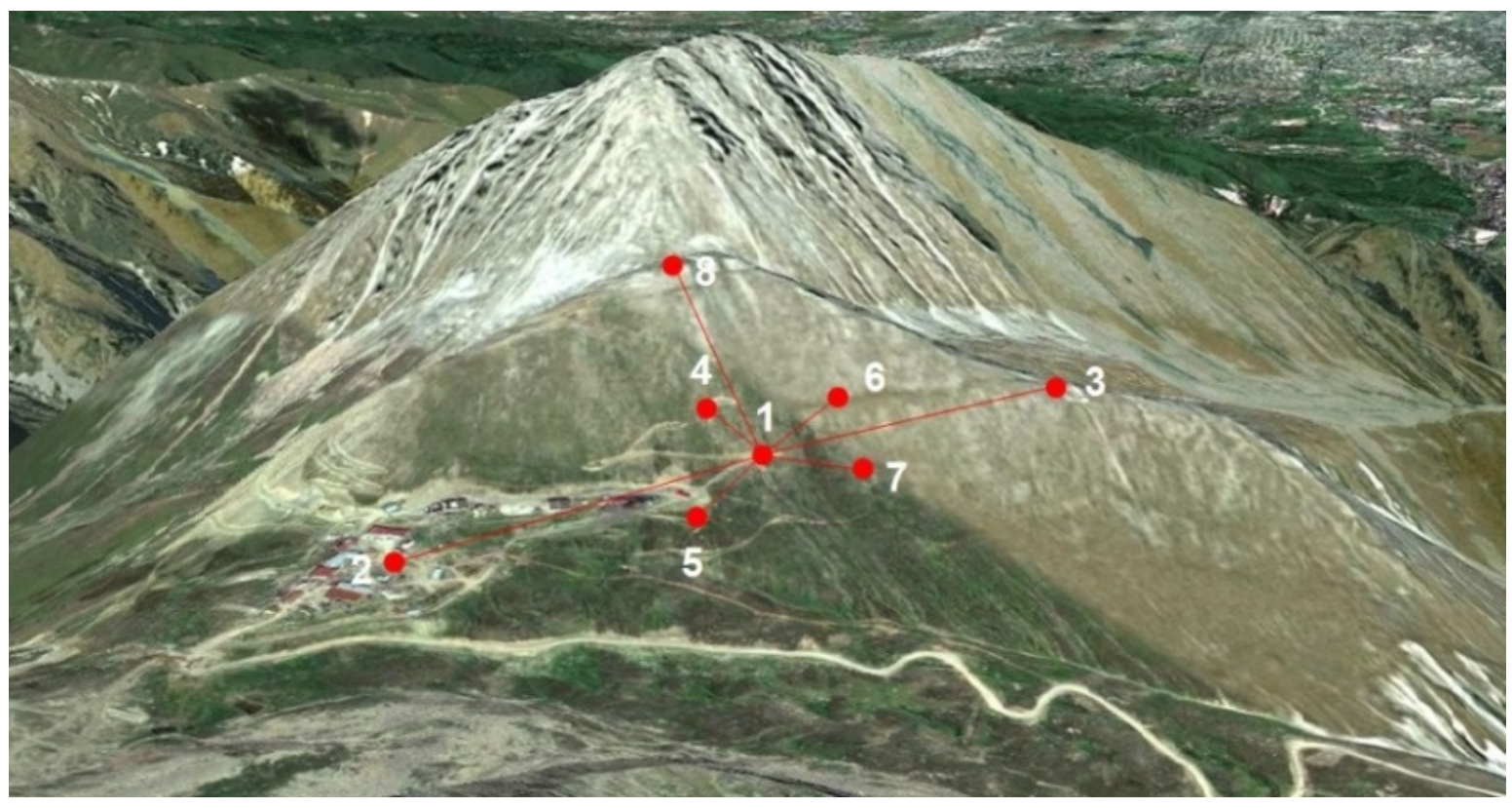

**Figure 1: The bird-eye view of the detector stations positions.**

**Table 1: Coordinates of detection points.**

| Station # | X, m | Y, m | Z, m | R, m |
|---|---|---|---|---|
| 1 | 0 | 0 | 0 | 0 |
| 2 | –445.9 | –85.6 | 2.8 | 454.1 |
| 3 | 384.9 | 79.5 | 36.1 | 394.7 |
| 4 | –55.0 | –94.0 | 31.1 | 113.3 |
| 5 | –142.4 | 36.9 | –12.6 | 147.6 |
| 6 | 151.2 | –17.9 | 31.3 | 155.4 |
| 7 | 88.6 | 178.4 | –39.0 | 203.0 |
| 8 | 221.3 | 262.0 | 160.7 | 378.7 |

The far periphery includes detection points 2, 3 and 8. Far points use only single SD each. They all have PMT-49 (FEU49) by MELZ [9] (replaced with R7723 in March 2017). This is due to the fact that long cables (~500m) are used to connect these points to the Data Acquisition system (DAQ) which is located at detection point 1. The cable calibration [3] shows that signal widening and signal loss become significant in this case, thus justifying the use of FEU49 as they have wider initial pulse. In the upgrade, they were all replaced as outdated.

All SD and GD are pyramid-shaped, the PMT above the center of detection medium at the height of the size of medium size (e.g. 1 m for scintillator and 0.5 m for glass).

For HT, the z-plane is parallel to the sky, the x-axis is directed north. All SD and GL are in the z-plane. There are also x- and y-plane scintillator detectors but they are not currently used. This arrangement is needed for the angular isotropy in the registration of charged particles to be used in the future. Theoretically, better isotropy may be achieved by an upgrade to liquid scintillators [10] [11] with a symmetric active volume but it is not being yet considered at this time.

## 3. Detectors MIP Response Calibration

Each SD/GD response to MIP is calibrated individually. For that, an additional trigger detector consisting of FEU49 and a 15 cm diameter scintillator is spaced under each detector being calibrated. Double-coincidence schema is used, facilitated by 14bit CAEN [12] DT5730 flash ADC. The setup schematic is shown in Figure 2. The reason for such setup design is that only two cables connect each detection point with DAQ physically.

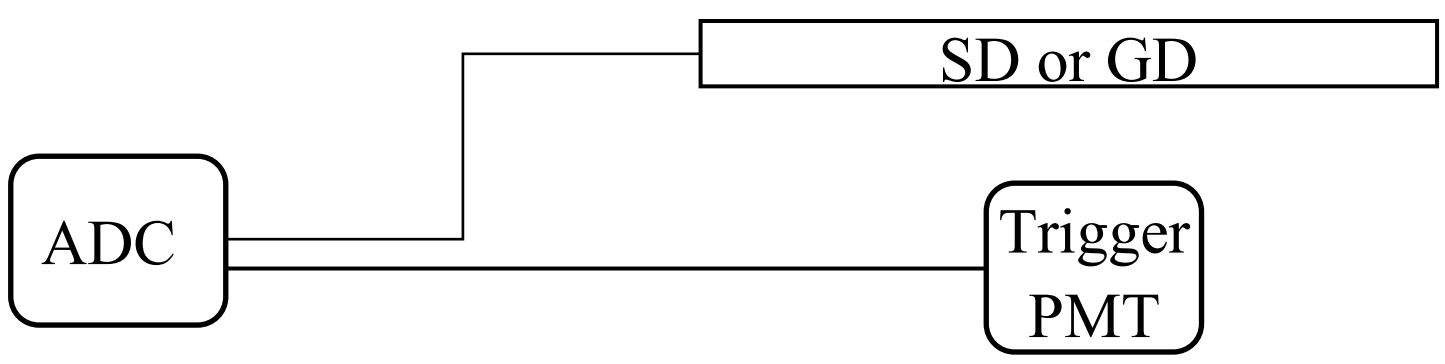

**Figure 2: MIP calibration setup schematic for SD/GD.**

The resulting calibration gives the area of a single MIP signal as well as the width of the MIP signal pulse from each SD. Due to data analysis details, total duration is currently taken as time between the 0.1 and 0.9 of the pulse area; pulse front is defined as time between 0.1 and 0.5 of the pulse area. This reduces the baseline noise effects. The uncertainty, associated with the size of the integration window is included in the total error.

From [1], the pulse front for SD is 7.16±0.40 ns and the total duration is 21.6±1.48 ns with the systematic error of ±0.10 ns; the pulse front for GD for one MIP is 2.17±0.13 ns (over 18 m cable) and the total duration is 5.10±0.67 ns with the ±0.11 ns systematic error for both.

The normalized area of a MIP signal from a GD from a detection point 1 is shown in Figure 3. The waveform recorded by the ADC consist of 5110 data points digitized every 2 ns each, for the total of 10.22 μs. The full range of $2^{14}$ bins corresponds to 2 V scale. The areas are given in the custom units of ADC counts · ns.

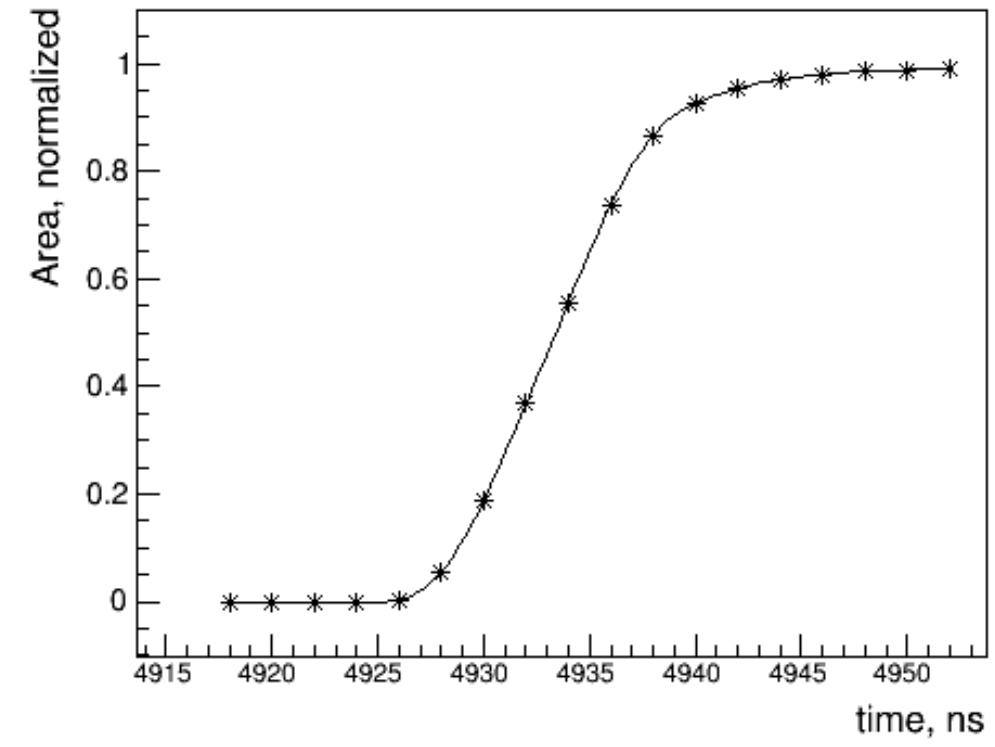

**Figure 3: Normalized integral for MIP pulse from GD.**

As with any calibration, where the detector that is being calibrated is also part of the triggering, there is also a question of the threshold enforced on the detector that is being calibrated. Here we use the threshold as low as possible but still above a pedestal value. For that, a threshold of a few mV is used first and the data is taken with a pedestal clearly visible. Then the data is retaken with the threshold value right above the pedestal value since for all detectors there is a clear separation between the pedestal and the lowest MIP signal. The Landau curve fit gives the most probable value (MPV) used for the calibration. Care is taken to make sure that chosen threshold value doesn't affect the resulting calibration value (e.g. any possible shift in area MPV is much smaller than the associated uncertainty). The example fit of a MIP signal area for the SD from detection point 1 (Center) is shown in Figure 4. Fit is done using ROOT [13] package PyROOT, areas are found using trapezoidal method [14]. The correspondence of the names of detection points to numbers is given in the

Table 2. A fit to a Landau distribution is used since the (relatively) thin target is used.

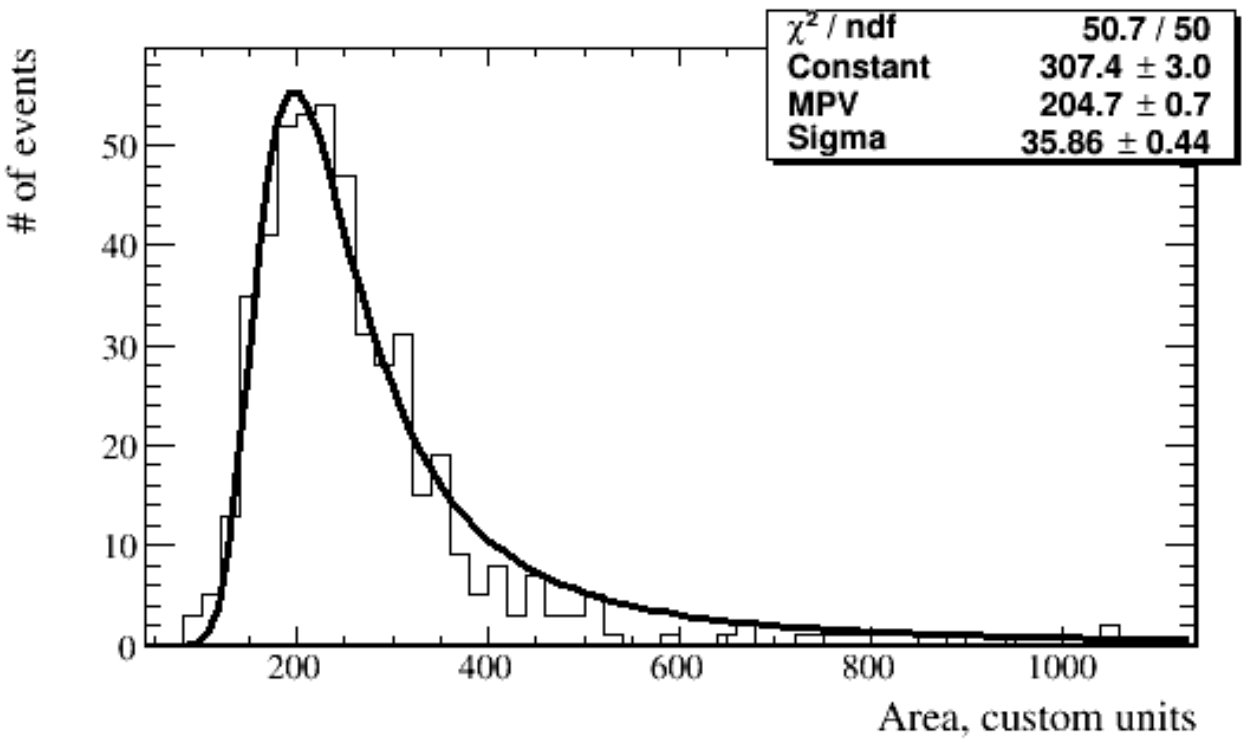

**Figure 4: SD detector response to MIP signal**

**Table 2: Names of detection stations**

| *Station name* | Center | Yastrebov | Stone Flower | Left | Kurashkin | Right | Bottom | Upper | Cher |
|---|---|---|---|---|---|---|---|---|---|
| *Station number* | 1 | 2 | 3 | 4 | 5 | 6 | 7 | 8 | VCD |

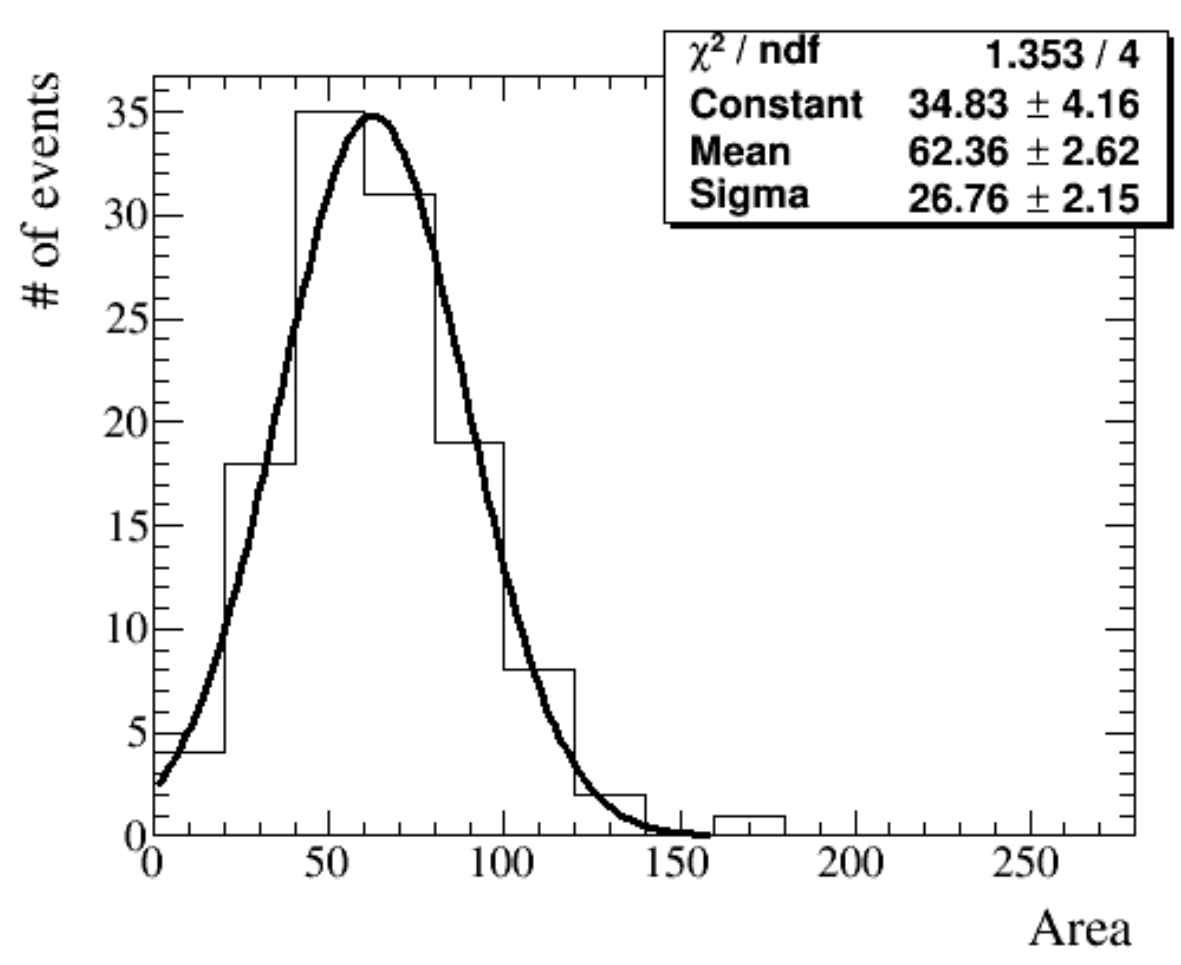

**Figure 5 : R7723 PMT single PE response pulse area at 1700V**

In order to obtain an approximate photon count for each MIP calibration for the match with simulation of the detectors we also do the single photo-electron (PE) R7723 PMT response pulse area calibration. For that, low light LED pulse is fed to PMT connected to DAQ via short cable (<1m). Since signal is baseline subtracted, the pedestal is very low and is removed from later fit. Care is taken to ensure that pedestal has about 80% of all events in order for single photon detection assurance. Figure 5 shows the pulse area of PMT single PE response at 1700V with pedestal subtracted. PMT is calibrated at the different bias voltages to match different detectors. Signal losses in each cable [3] and presence of impedance matching resistor (if present) must be accounted for when number of photons is accessed for each detector.

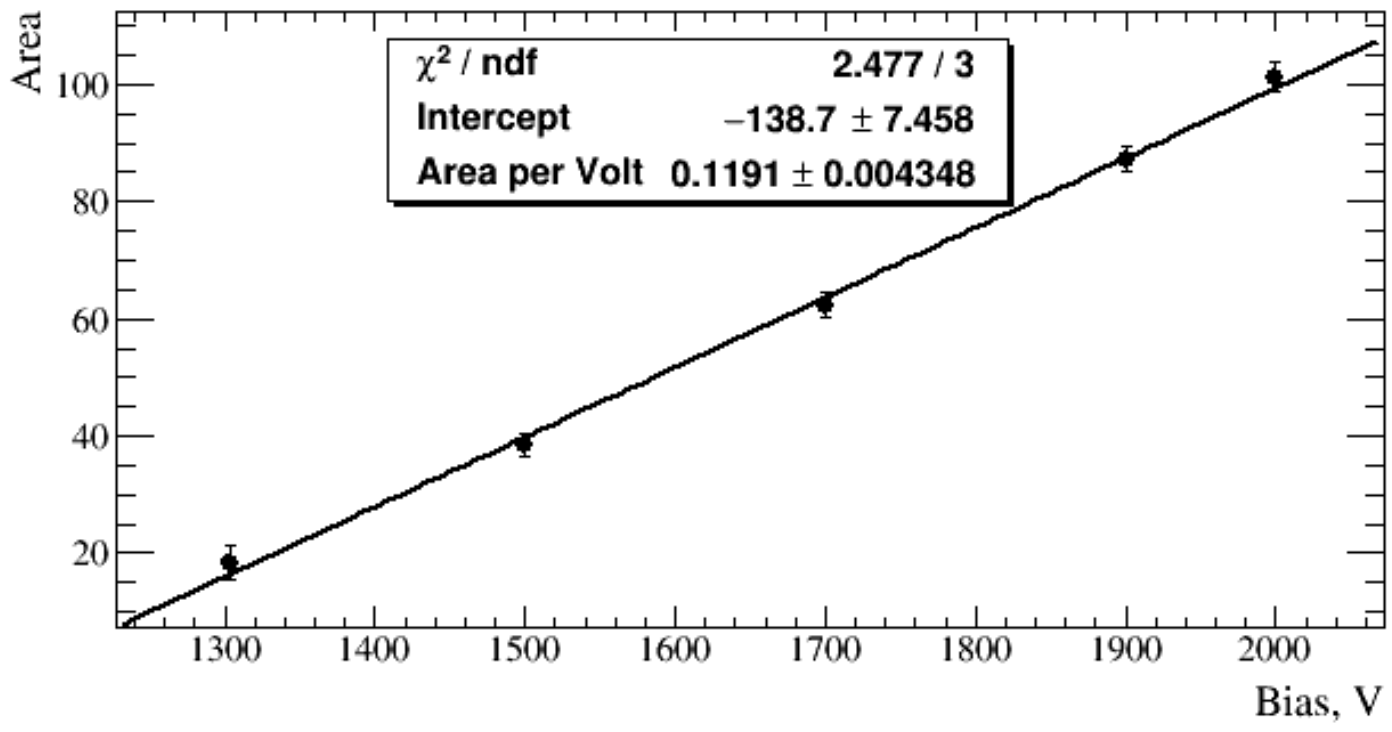

**Figure 6: Single PE pulse area vs. bias voltage for R7723 PMT**

**Table 3: Detector MIP response pulse area at operating bias voltage**

| Detection point and cable designation | Detector type | Area (ADC counts · ns) | | Detection point and cable designation | Detector type | Area (ADC counts · ns) |
|---|---|---|---|---|---|---|
| Bottom New | SC | 529 ± 10<br>91 ± 7 | | Left New | SC | 574 ± 11<br>110 ± 8 |
| Bottom Old | GL | 134 ± 3<br>23 ± 2 | | Left Old | GL | 132 ± 2<br>16 ± 2 |
| Center Blue | Empty | | | Right New | SC | 513 ± 8<br>78 ± 5 |
| Center Green | Empty | | | Right Old | GL | 137 ± 3<br>31 ± 2 |
| Center Red (before Oct. 20, 2016) | GL | 83 ± 3<br>22 ± 2 | | Stone Flower New | SC/FEU49 | 710 ± 15<br>201 ± 10 |
| Center Red New | GL | 85 ± 3<br>23 ± 2 | | Stone Flower Old | Empty | |
| Center White | Empty | | | Yastrebov New | SC/FEU49 | 2110 ± 16<br>429 ± 11 |
| Center Yellow | SC | 470 ± 5<br>57 ± 6 | | Yastrebov Old | Empty | |
| Cher Green Red | Empty | | | Upper New | SC/FEU49 | 768 ± 6<br>124 ± 7 |
| Cher White Blue | VCD | | | Kurashkin New | SC | 549 ± 12<br>107 ± 8 |
| Cher Yellow | VCD | | | Kurashkin Old | GL | 155 ± 5<br>50 ± 3 |
| Center Blue New | GL | 183 ± 9<br>44 ± 5 | | Bottom Old2 | Empty | |

In Figure 6, the single PE area vs biasing voltage is shown. For the R7723 PMT, we can see that it is linear for a very wide range of biases, with 2000V being recommended maximum and pulses becoming too close to noise floor below 1300V for accurate calibration.

The results of the MIP calibration are in Table 3. Only values at used operating biases are shown per detector per cable. The full calibration table is in Appendix I. All cable effects are included in the calibration; MPV (top value) and σ (bottom value) with corresponding fit uncertainties are listed.

## 4. SD Detector Response Uniformity

For SD detectors, due to the PMT placement above the scintillator center within pyramid-shaped enclosure, the non-uniformity of particle detection from scintillator center and edges exists. In order to accurately assess the charged particles flux through each SD, this non-uniformity should be measured.

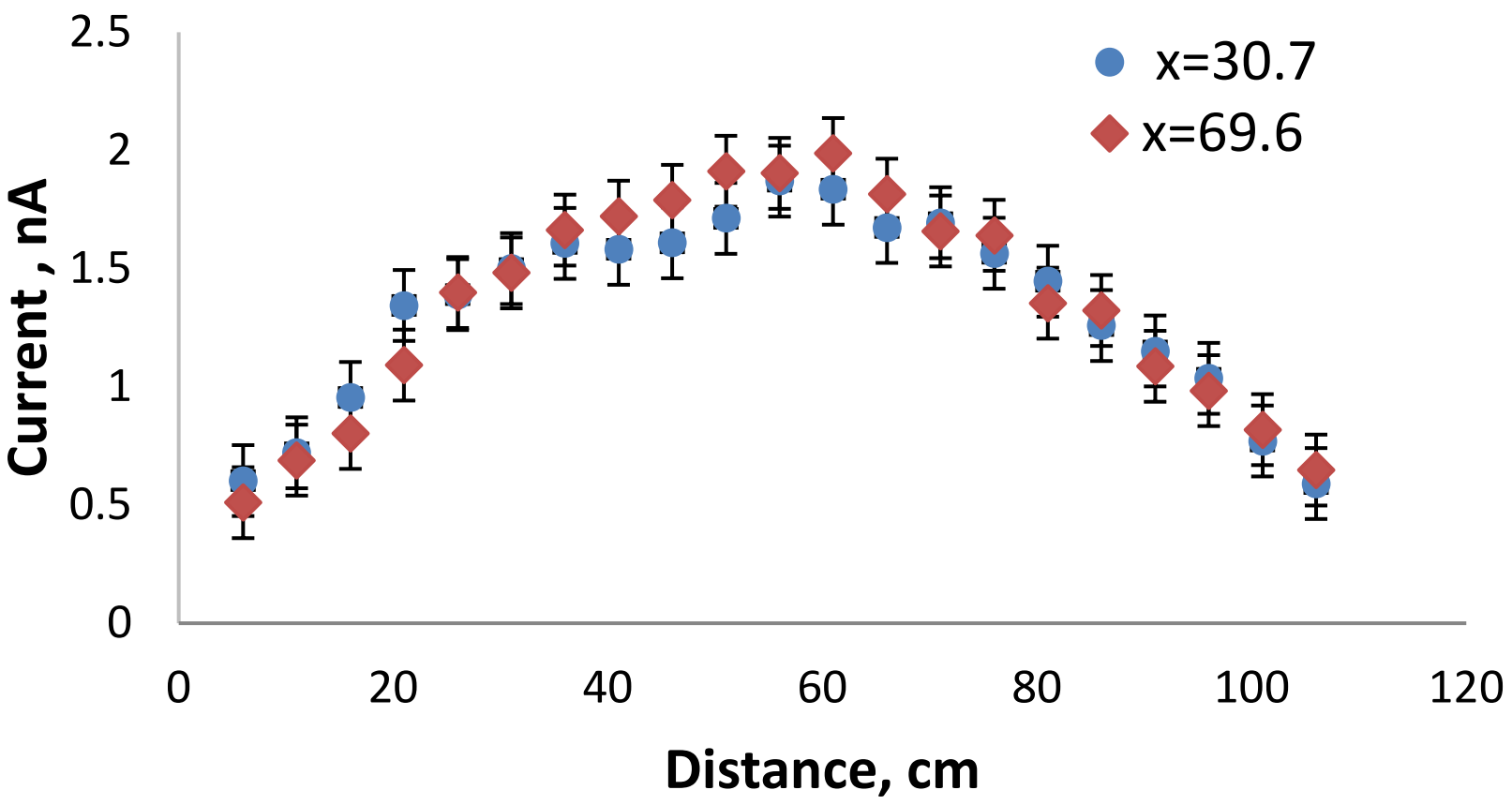

For this purpose, each SD is scanned using $^{60}$Co radioactive source across the scintillator side along four lines: both diagonals and two lines, passing though centers of parallel sides in x and in y in scintillator plane. From the data, an average weight of 0.70±0.06 is calculated that is later applied to MIP calibration for each SD detector. The GD uniformity is 0.74±0.04 [7].

## 5. Cherenkov Detector

The Vavilov-Cherenkov Detector (VCD) is located next to detection point 1 as close to DAQ system as possible. The VCD consists of three parabolic mirrors of 150cm diameter with 65cm focal length each mounted on the rotating support allowing registration of radiation in zenith angle range of 0°-80° and azimuthal angle range of 0° -360°. There is a PMT-49 (FEU49B) and a Hamamatsu H6527 PMT located in the focal point of two lower mirrors. Both are 15cm cathode diameter PMTs with the spectral response from 360 nm to 600 nm. The angle of view for each mirror is ~13°. For completeness, we mention the VCD but its calibration is not included here. Note that PMTs are very easily damaged by the light, thus a future upgrade may include the Geiger-mode avalanche photodetector [15] [16] arrays [17] since they are unaffected by high light intensity such as moonlight or car headlights. These are fast detectors and have been used on a large scale [18]. Both VCD channels are connected to DAQ using separate cable each.

## 6. Conclusion

The MIP calibration of the HT detector system was carried out for both SD and GL detectors at different biasing voltages. The uniformity of response for these detectors was measured as well. A detector upgrade is planned so this note will not be updated and will be valid for data before March 2017

## Appendix I

Calibrations valid before March 2017. Updated version will be uploaded as a separate document as some detectors were changed in HT

MIP calibration for all scintillator detectors at different bias voltages. In green is the current operation bias.

| Name | Voltage | MPV | MPV Error | sigma | Sigma Error |
|---|---|---|---|---|---|
| Bottom SC | 1100 | 218.4 | 4.5 | 44.2 | 2.5 |
| | 1200 | 342.7 | 7.3 | 60.8 | 3.7 |
| | 1250 | 418.1 | 8 | 76.5 | 4.9 |
| | 1300 | 528.6 | 10 | 90.9 | 6.6 |
| | 1350 | 629.7 | 13.8 | 109.8 | 8.7 |
| | 1400 | 694.3 | 13.4 | 123.2 | 8.3 |
| Kurashkin SC | 1200 | 412.2 | 7.1 | 68.9 | 4.6 |
| | 1250 | 462.1 | 7.5 | 80.8 | 5.1 |
| | 1300 | 549.1 | 12.3 | 106.9 | 7.6 |
| | 1400 | 765.1 | 14.6 | 138.4 | 11.2 |
| Left SC | 1100 | 370.7 | 2.8 | 61.4 | 1.6 |
| | 1200 | 488.1 | 7.5 | 81.9 | 4.7 |
| | 1250 | 573.8 | 11 | 109.8 | 7.7 |
| | 1300 | 678.7 | 14 | 125.3 | 8.8 |
| | 1400 | 929.7 | 23 | 212.4 | 16.1 |
| Right SC | 1200 | 396.5 | 5.1 | 50.9 | 2.7 |
| | 1300 | 512.7 | 7.9 | 78.3 | 4.6 |
| | 1350 | 602.1 | 9.2 | 93.2 | 5.7 |
| | 1400 | 687.4 | 10.2 | 110.5 | 6.1 |
| | 1500 | 900.5 | 15.3 | 160.8 | 9.8 |
| St Flower SC | 1862 | 710 | 15 | 201 | 10 |
| Upper SC | 1700 | 768 | 6 | 124 | 7 |
| Yastrebov SC | 1800 | 615 | 10 | 221 | 4 |
| | 1900 | 695 | 12 | 253 | 5 |
| | 2000 | 1050 | 12 | 300 | 9 |
| | 2100 | 1420 | 25 | 370 | 13 |
| | 2200 | 2110 | 16 | 429 | 11 |
| | 2300 | 2695 | 15 | 536 | 11 |
| Center SC | 1000 | 200 | 4 | 37.3 | 0.5 |
| | 1100 | 260 | 5 | 35.2 | 2.2 |
| | 1200 | 470 | 5 | 56.8 | 6 |
| | 1500 | 1160 | 6 | 205 | 6 |
| Bottom Gl | 1600 | 57.9 | 2.4 | 22.9 | 1.5 |
| | 1700 | 70.5 | 2 | 20.2 | 1.2 |
| | 1800 | 91.3 | 2.1 | 22.2 | 1.2 |
| | 1900 | 113 | 2.2 | 23.1 | 1.5 |
| | 2000 | 134 | 2.6 | 23.2 | 1.5 |
| | 2100 | 148 | 3 | 27.9 | 1.9 |
| Left Gl | 1700 | 99.7 | 2.5 | 17.7 | 1.8 |
| | 1800 | 115 | 2.4 | 19.9 | 1.8 |
| | 1900 | 132 | 2.1 | 16 | 1.6 |
| | 2000 | 157.5 | 3.3 | 26.3 | 2.24 |
| Right Gl | 1800 | 64.2 | 2.3 | 20.9 | 1.6 |
| | 1900 | 95.1 | 2.8 | 27.3 | 1.8 |
| | 2000 | 136.7 | 3.2 | 30.9 | 2 |
| | 2100 | 173.4 | 3.5 | 34.2 | 2.2 |
| Center GlBig | 1400 | 137.1 | 2.2 | 26.9 | 1.3 |
| | 1500 | 182.6 | 8.5 | 43.6 | 5.2 |
| | 1600 | 247.5 | 5.6 | 43.1 | 3.9 |
| | 1700 | 422 | 9 | 79 | 6 |
| Kurashkin Gl | 1800 | 91.2 | 3.6 | 22.9 | 2.2 |
| | 1900 | 155.4 | 4.8 | 50.2 | 2.9 |
| | 2000 | 214.9 | 6.7 | 60.3 | 3.5 |
| | 2100 | 274.1 | 23.6 | 75.8 | 8.4 |
| Center Gl2cm | 1700 | 61.7 | 2.2 | 20.1 | 1.1 |
| | 1800 | 68.9 | 2 | 18.9 | 1.2 |
| | 1900 | 76.4 | 1.7 | 15.6 | 1.2 |
| | 2000 | 84.5 | 2.4 | 22.7 | 1.4 |
| Center Gl3cm | 1900 | 83.1 | 2.4 | 22.1 | 1.4 |